\documentclass[12pt]{article}
\usepackage{amsmath,amsfonts}
\usepackage{latexsym,amsmath,amsfonts,amssymb}
\usepackage[latin1]{inputenc}
\usepackage[pdftex]{graphicx}
\usepackage{epsfig}

\usepackage{euscript,mathrsfs}
\usepackage{fullpage}
\usepackage{color}
\usepackage{bbm}
\usepackage[american]{babel}

\makeatletter\@addtoreset{equation}{section}\makeatother

\def\be{\begin{equation}}
\def\ee{\end{equation}}
\def\bea{\begin{eqnarray}}
\def\eea{\end{eqnarray}}

\newcommand{\p}{\partial}

\newcommand{\nn}{{\nonumber}}

\renewcommand{\title}[1]{\vbox{\center\LARGE{#1}}\vspace{5mm}}
\renewcommand{\author}[1]{\vbox{\center#1}\vspace{5mm}}
\newcommand{\address}[1]{\vbox{\center\em#1}}
\newcommand{\email}[1]{\vbox{\center\tt#1}\vspace{5mm}}

\begin{document}
\begin{titlepage}
\begin{center}
\hfill \\
\hfill \\
\vskip 1cm

\title{Holographic Symmetry-Breaking Phases\\ in AdS$_3$/CFT$_2$}

\author{Nima Lashkari}

\address{McGill Physics Department, 3600 rue University,\\
Montr{\'e}al, QC H3A 2T8, Canada}

\email{lashkari@physics.mcgill.ca}

\end{center}

\abstract{
In this note we study the symmetry-breaking phases of 3D gravity coupled to matter. In particular, we consider black holes with
scalar hair as a model of symmetry-breaking phases of a strongly coupled 1+1 dimensional CFT. In the case of a discrete symmetry, we show that these theories admit
phases of broken symmetry and study the thermodynamics of these phases. We also demonstrate that the 3D Einstein-Maxwell theory shows continuous symmetry breaking
at low temperature. The apparent contradiction with the Coleman-Mermin-Wagner theorem is discussed.}

\vfill

\end{titlepage}

\eject \tableofcontents
\section{Introduction}
In light of recent developments, gauge-gravity dualities \cite{Maldacena:1997re,Witten:1998qj} have proven to go beyond the early examples of conformal gauge theories and
could potentially be applied to many strongly coupled condensed matter systems as well as non-conformal gauge theories. 
% This is of particular interest
% for condensed matter physics since the regime where gauge-gravity techniques become relevant suffers from a failure of standard analytic tools due to the strong coupling dynamics.
Gravity models dual to condensed matter systems close to quantum critical points have been constructed and extensively studied in the literature. For a review see \cite{Hartnoll:2009sz,McGreevy:2009xe}.

Symmetry breaking and phase transitions in strongly coupled media are examples where one might hope to obtain insight into the system by studying its gravity dual.
%  \cite{Klebanov:1999tb}. 
This correspondence has also been used in the opposite direction to shed light on gravity by using the intuition provided to us by the tools of quantum field theories. 
For instance, the observation that low temperature order is commonplace in field theories has led to the discovery of violations of the no-hair theorem in asymptotically $AdS$ space-times \cite{Gubser:2000ec,Gubser:2000mm}.
Indeed, it turns out that there is a landscape of gravity-matter theories with a rich phase space of hairy black holes at low temperature \cite{Denef:2009tp}. Phase transitions of this type in 4D charged black holes coupled to scalars have been analyzed in detail
 as a valuable phenomenlogical tool for 2+1 dimensional superconductors \cite{Gubser:2005ih,Hartnoll:2008vx,Horowitz:2008bn,Hartnoll:2008kx}.

The purpose of this work is to generalize the examples of holographic symmetry breaking to 3D matter-gravity theories. Working in three dimensions has the benefit that the dynamics are much simpler and one can hope for better analytic and numerical control on the physics of these holographic phases. 

Three dimensional gravity with a negative cosmological constant admits topological black hole solutions denoted as BTZ \cite{Banados:1992wn}. In the presence of matter however, the theory is no longer topological and hairy black holes with the standard asymptotically $AdS_3$ boundary conditions
% are not necessarily locally $AdS_3$. There is a vast literature in this fiel \cite{Carlip:1995qv}. The conformally coupled scalar hair for three dimensional black holes was first found in \cite{Martinez:1996gn}. In \cite{Henneaux:2002wm}, the authors constructed the first example of a hairy black hole
with a minimally coupled scalar in 3D. Gravity coupled to a $U(1)$ gauge field has charged black holes solutions
that are commonly referred to as charged BTZ's \cite{Martinez:1999qi}, and in \cite{Maity:2009zz} charged BTZ was used as a holographic dual to Fermi-Luttinger liquids. 
% Similar holographic models of quantum Fermi liquids were studied in \cite{David:2009np,Hung:2009qk}. In the extremal limit, the near horizon geometry of these black holes has an $AdS_2$ factor. From the point of view of the dual field theory, 
% these extremal geometries describe a flow between two 2D conformal field theories in UV and IR \cite{Hotta:2008xt,Cadoni:2008mw}. 

% According to the standard AdS/CFT dictionary, working in the classical gravity regime translates to a strongly coupled dual field theory with very large central charge. Matter field condensates near the black hole horizon are dual to the expectation values of operators' one-point functions in the field theory. The expectation value of the field theory operator 
% corresponding to a bulk scalar field can be taken to be the order parameter for the phase transition dual to the condensation of the scalar field near the black hole horizon.

In section 2, we start with a review of an analytic example of a hairy three dimensional black hole. Then, we add an extra interacting scalar field to the
model to obtain a phase transition. We find a tower of symmetry-breaking phases that appears infinite. Surprisingly, we observe that these symmetry breaking phases survive at large temperatures.
The thermodynamic properties of these states are studied in detail.

We briefly discuss the  Coleman-Mermin-Wagner theorem in section 3 and give an example of continuous symmetry breaking in three dimensions. Section 4 is devoted to a discussion of the implications of
our results and other interesting extensions of this work.

As we were in the process of completing this work we learned about the work in \cite{Ren:2010ha} that has some overlap with the material presented here in section \ref{cont}.
\section{Discrete Symmetries}

\subsection{An Analytic Scalar Hair}
An analytic solution of scalar gravity with a one parameter family of potentials for the scalar was obtained in \cite{Henneaux:2002wm}. The thermodynamics of
these black holes was studied in \cite{Gegenberg:2003jr}. We briefly review these results here. 

Consider the action
\begin{equation}
 S=\frac{1}{16\pi G}\int d^3x \sqrt{-g}\left(R-\frac{1}{2}(\p\phi)^2-V(\phi)\right),
\end{equation}
with $V(\phi)=-\frac{2}{l^2}\left(\cosh^6(\phi/4)+\nu \sinh^6(\phi/4)\right)$ where $\nu\ge -1$ could be interpreted as a self-coupling in the conformal frame \cite{Henneaux:2002wm}.
This potential has a maximum at $\phi=0$. We are interested in normalisable solutions where $\phi$ vanishes only at $r\to\infty$. This provides us with the asymptotically locally
AdS space-time we need for holography.

The solution with the scalar hair is
\begin{eqnarray}\label{hairyanalytic}
\phi(r)&=&4\: \text{arctanh}\:\sqrt{\frac{B}{H+B}},\nn\\
 ds^2&=&-\left(\frac{H}{H+B}\right)^2F(r)dt^2+\left(\frac{H+B}{H+2B}\right)^2\frac{dr^2}{F(r)}+r^2d\theta^2,\nn
\end{eqnarray}
with
\begin{eqnarray}
H(r)&=&\frac{1}{2}\left(r+\sqrt{r^2+4Br}\right),\nn\\
F(r)&=&\frac{H^2}{l^2}-(1+\nu)\left(\frac{3B^2}{l^2}+\frac{2B^3}{l^2H}\right),
\end{eqnarray}
and $B\geq 0$ is the only integration constant. An important feature of this solution is that one can not switch off the scalar field without changing the mass of the black hole.

\begin{figure}[h]
\centerline{a) \epsfig{figure=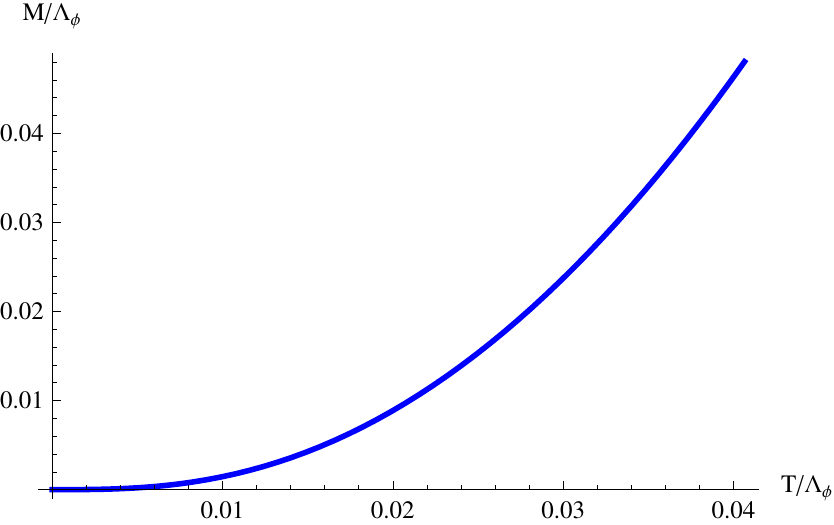, width=3in}  b) \epsfig{figure=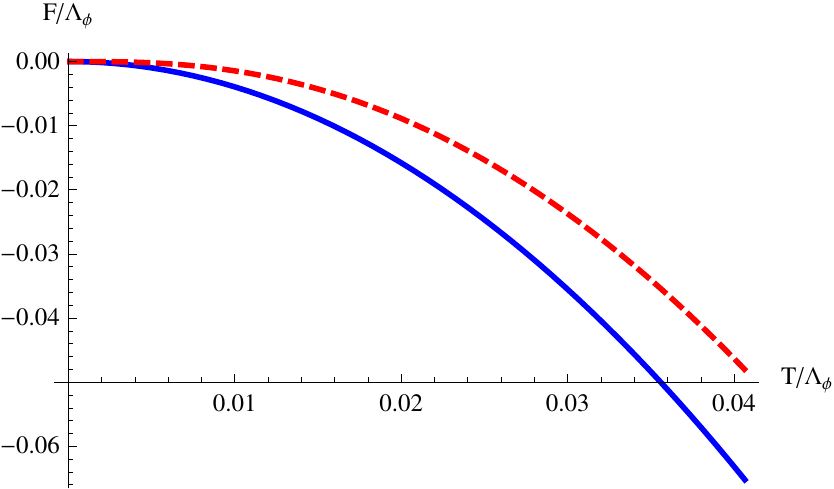, width=3in}}
\caption{
\label{analytic}
The thermodynamic properties of hairy black holes: a) the mass of the hairy solution b) at any temperature, the free energy of BTZ black holes (blue solid line) is less than the 
black hole with the scalar condensate (red dashed line).}
\end{figure}

According to AdS/CFT, the radial profile of the scalar field in the bulk is 
dual to the renormalization group running of an operator $\mathcal{O}_\phi$ in the field theory. Here, the constant $B$ controls the scale where the relevant deformation $\Lambda_\phi$ becomes important in the dual field theory.
This scale can be read from the near
infinity expansion of the field $\phi$. Setting $\Lambda_\phi=1$ amounts to $B=1/16$. In these units, the horizon is at $r_0= \Theta_\nu/16$ with $\Theta_\nu$ defined by
\begin{equation}
 \Theta_\nu=2(\bar{z}z)^{2/3}\left(\frac{z^{2/3}-\bar{z}^{2/3}}{z-\bar{z}}\right),\qquad z=1+i\sqrt{\nu}.
\end{equation}
At any given temperature, the action admits two solutions, one with scalar hair in $\ref{hairyanalytic}$ and the other one the non-rotating BTZ
black holes with $\phi=0$ everywhere in the bulk. The hairy phase in (\ref{hairyanalytic}) breaks the $\mathbb{Z}_2$ symmetry of $\phi\to -\phi$ in the action. The thermodynamic quantities
 are found from the smoothness of the euclidean geometry on the horizon and the near infinity expansion of fields. In units where $8G=1$ and $l=1$ one finds
\begin{eqnarray}
&& T=\frac{3(1+\nu)}{32\pi\Theta_\nu},\qquad S=\frac{A}{4G}=\frac{\pi\Theta_\nu}{4}\nn\\
&& F=-M=-\frac{3(1+\nu)}{256}.
\end{eqnarray}
It turns out that BTZ black holes of the same temperature have lower free energy than hairy solutions and win the thermodynamic competition , see figure \ref{analytic}. 
The hairy solutions describe $\mathbb{Z}_2$ broken phases with $\langle O_\phi\rangle=-1/24$. In the zero temperature limit of (\ref{hairyanalytic}), the horizon area 
vanishes, joining the BTZ branch at the zero mass black holes. This was physically expected since a finite area at zero temperature
corresponds to a highly degenerate ground state in the dual field theory \cite{Horowitz:2009ij}.

\subsection{High Temperature Order}
The example of the hairy black hole discussed in the previous section has the benefit of being analytic. However, due to the simplicity of the model, 
the value of the condensate is independent of temperature. In other words, there is no phase transition between the BTZ and the scalar condensate
at finite temperature. In order to obtain a temperature dependent order parameter, one needs to add more ingredients to the model.
A five-dimensional theory with two interacting scalars was introduced in \cite{Gubser:2005ih} which exhibits a phase transition between a symmetry-breaking phase and the
symmetric phase. A four-dimensional analogue of this system has been studied in \cite{Buchel:2009ge}. In this section, we will discuss a three-dimensional model
of this kind in detail.

Let us start with two massive scalar fields in three-dimensional gravity coupled according to:
\begin{eqnarray}
&& S=\frac{1}{16\pi G}\int d^3x\sqrt{-g}\left(R-\frac{1}{2}(\p\phi)^2-\frac{1}{2}(\p\chi)^2-V(\phi,\chi)\right),\nn\\
&& V(\phi,\chi)=-\frac{2}{l^2}+\frac{1}{2}m_\phi^2\phi^2+\frac{1}{2}m_\chi^2\chi^2-\frac{\lambda^2}{2}\phi^2\chi^2,
\end{eqnarray}
where $l$ is the length scale of the $AdS_3$ vacuum. Scalar masses are chosen to be $-1< (m_\phi l)^2<0$ and $0<(m_\chi l)^2$. The mass of the scalar $\phi$ is in the range
that allows two different quantizations according to the AdS/CFT dictionary \cite{Klebanov:1999tb}. We follow the standard quantization for the field $\phi$ by
fixing the dominant mode at infinity. The idea is that we turn off $\chi$ and solve for hairy
black holes with scalar $\phi$. As we make the condesate of $\phi$ larger near the horizon, the field $\chi$ becomes turned on through the interaction
term $\phi^2\chi^2$ breaking the $\chi\to -\chi$ symmetry of the theory. The aim is to study these symmetry-breaking solutions.

\subsubsection{Probe limit}
It is heuristic to take a probe limit where matter fields propagate on a rotating BTZ geometry.
Scaling $\phi\to \phi/\lambda$ and $\chi\to\chi/\lambda$, the matter stress tensor transforms as $T_{\mu\nu}\to T_{\mu\nu}/\lambda^{2}$.
By sending $\lambda\to \infty$ one can make sure that the back-reaction of scalars on the geometry is negligible. In this limit, the solutions of
the Einstein equation are BTZ black holes. The relevant equations for the propagation of scalars on a BTZ geometry are
\begin{equation}\label{prop}
 \Box\phi=\phi(m_\phi^2-\chi^2),\qquad \Box\chi=\chi(m_\chi^2-\phi^2).
\end{equation}
In the absence of the $\chi$ field, there are two independent solutions to the equation of motion of $\phi$:
\begin{eqnarray}\label{wave}
 \phi^\pm(u)=u^{\:-\Delta_\pm/2}{}_2F_1\left[\Delta_\pm/2,\Delta_\pm/2,\Delta_\pm,\frac{1}{u}\right],
\end{eqnarray}
where
\begin{equation}\label{delta}
 u=\frac{r^2-r_-^2}{r_+^2-r_-^2},\qquad \Delta_\pm=1\pm\sqrt{1+(l m_\phi)^2}.
\end{equation}
%  Clearly near the boundary $u\to\infty$, the solution 
% falls with exponents $h_+$ and $h_-$. 
For scalars with mass $-1\leq (m_\phi l)^2\leq 0$ both of the modes $\phi^\pm$ are normalizable. This solution is
a monotonically decreasing function of $r$ with its maximum value on the horizon $\phi_0$. 

In (\ref{prop}) the field $\chi$ has an effective mass term $m_\chi^2-\phi^2$.
When $\phi_0$ becomes comparable to $1+(m_\chi l)^2$ in (\ref{delta}), $\Delta_\pm$ becomes imaginary and the field $\chi$ violates the BF bound near the horizon and condenses, breaking the initial
$\mathbb{Z}_2$ symmetry, $\chi\to -\chi$. Note that as it is clear from (\ref{wave}) for imaginary $\Delta_\pm$, the solution has an oscillatory behavior.  
% As long as the effective mass is positive, 
% in (\ref{prop}) $\chi''$ is proportional to $\chi$ on local extrema and therefore
% $\chi(r)$ is monotonic. However, once we make $\phi$ large enough this is no longer the case and $\chi$ can have nodes. 
It turns out that as we increase $\phi_0$ there is a 
discrete set of solutions for $\chi$ with zeros at finite $u$. These symmetry-breaking solutions with higher number of nodes and their thermodynamic properties are the focus of this section.
\subsubsection{Holographic renormalization}
Now we turn to the fully back-reacted equations of motion. We consider the following ansatz for the metric
\begin{equation}
 ds^2=-f(r)dt^2+\frac{dr^2}{h(r)}+r^2(d\theta+g(r)dt)^2.
\end{equation}
There are two equations for the dynamics of scalar fields:
\begin{eqnarray}\label{scalardyn}
 &&\phi''+\phi'\left(\frac{1}{r}+\frac{f'}{2f}+\frac{h'}{2h}\right)-\frac{1}{h}\frac{\p V}{\p\phi}=0\nn\\
&&\chi''+\chi'\left(\frac{1}{r}+\frac{f'}{2f}+\frac{h'}{2h}\right)-\frac{1}{h}\frac{\p V}{\p\chi}=0,
\end{eqnarray}
and four Einstein's equations out of which only three are independent:
\begin{eqnarray}\label{backreac}
&&\frac{-3}{r}+\frac{f'}{2f}-\frac{h'}{2h}-\frac{g''}{g'}=0\nn\\
&& \frac{2f''}{f}-\frac{f'^2}{f^2}-\frac{4r^2g'^2}{f}-\frac{2h'}{r h}+\frac{f'h'}{f h}=0\nn\\
&&\frac{2V(\phi,\chi)}{h}+\frac{f'}{rf}+\frac{h'}{rh}+\frac{r^2g'^2}{f}=0.
%&& \phi'^2+\chi'^2-\frac{f'}{rf}+\frac{h'}{rh}=0\nn\\
\end{eqnarray}
Here primes denote derivatives with respect to $r$. 
The first two equations in (\ref{backreac}) could be integrated once to give respectively
\begin{eqnarray}\label{g'}
 &&g'(r)=\sqrt{\frac{f}{h}}\frac{c}{r^3},\nn\\
&&r\sqrt{\frac{h}{f}}f'-2\sqrt{f h}-2 c g=\lambda.
% &&\int dr\:\left(r\:V\:\sqrt{\frac{f}{h}}\right)+\sqrt{fh}+\frac{c g}{2}=0,
\end{eqnarray}
where $c$ and $\lambda$ are constants of integration. We will see later that $c$ is the angular momentum of the black hole. Notice that due to the reparametrizations, $f$  and $g$ are determined only up to the multiplication and addition of a constant, respectively.  
From the periodicity in the Euclidean time, the temperature of this geometry could be read off to be  
\begin{equation}
 T=\frac{1}{4\pi}\left(\sqrt{\frac{h}{f}}f'\right)\Big|^{r=r_0},
\end{equation}
where $r_0$ is the location of the outer horizon.
The expansion of fields near the boundary depends on the value of $m_\phi^2$. For $-1< (m_\phi l)^2<-3/4$ and $0<(m_\chi l)^2$, solving the equations of motion order by order in $1/r$ gives
\begin{eqnarray}
&& \phi(r)=\frac{\phi_0}{r^{(2-\Delta_\phi)}}+\frac{\phi_1}{r^{\Delta_\phi}}+O\left(r^{p\leq 3-\Delta_\phi}\right),\qquad 
\chi(r)=\frac{\chi_0}{r^{\Delta_\chi}}+O\left(r^{p<1-\Delta_\chi}\right),\nn\\
&& f(r)=f_0 r^2-\frac{\sqrt{f_0}\lambda}{2}+O\left(r^{p< 0}\right),\nn\\
&& h(r)=r^2+\frac{\phi_0^2(2-\Delta_\phi)}{2}\:r^{2(\Delta_\phi-1)}+\left(\phi_0\phi_1\Delta_\phi(2-\Delta_\phi)-\frac{\lambda}{2\sqrt{f_0}}\right)+O\left(r^{p\leq 0}\right),\nn
\end{eqnarray}
where we have suppressed indices by using the notation $\Delta_\phi=\Delta_{+,\phi}$. The function $g(r)$ is found from (\ref{g'}).

The Euclidean action could be written as
\begin{equation}
 I_E=\frac{-1}{16\pi G}\int d\tau d\theta dr\sqrt{-g}\left(R-\frac{(\p\phi)^2}{2}-\frac{(\p\chi)^2}{2}-V\right)-\frac{1}{8\pi G}\int_{\p\mathcal{M}} d\tau d\theta \sqrt{-\gamma}K+I_{c.t.}.
\end{equation}
The second term is the boundary Gibbons-Hawking term; $\gamma$ is the induced metric on the boundary and $K$ is the extrinsic curvature on $r=const.$ hypersurface. 
Using Einstein's equations the bulk action is evaluated on-shell to be
\begin{equation}
 I^{reg}_{bulk}=-\frac{\beta}{4G}\int_{r_0}^{1/\epsilon} dr\: \sqrt{\frac{f}{h}}\: r V(\phi,\chi)=\frac{\beta}{4G}\left[\sqrt{fh}\: \Big|^{1/\epsilon}+\frac{c g}{2}\:\Big|^{1/\epsilon}_{r_0}\right],
\end{equation}
where we have used the third equation in (\ref{backreac}). The integral is regulated by introducing the IR cut-off at $r=1/\epsilon$. 
The Gibbons-Hawking term is 
\begin{equation}
 I_{G.H.}=-\frac{\beta}{4G}\left[\sqrt{fh}+\frac{rf'}{2}\sqrt{\frac{h}{f}}\right].
\end{equation}
Then, the full Euclidean action is found to be
\begin{equation}
 I_{E}=-\frac{\beta}{8 G}\left[c g\Big|^{r_0}+r f' \sqrt{\frac{h}{f}}\:\Big|^{1/\epsilon}\right]+I_{c.t.}.
\end{equation}
The infinities in the on-shell action could be removed by adding the local counter term\footnote{Holographic renormalization and this counter-term were first introduced in \cite{de Haro:2000xn}.}
\begin{equation}
 I_{c.t.}=\frac{1}{8\pi G}\int d\tau d\theta \sqrt{-\gamma}\left[1+\frac{\Delta_\phi^-\phi^2}{4}\right].
\end{equation}
Note that the scalar field $\chi$ falls so fast that the bulk integral of its on-shell action is convergent.

According to the AdS/CFT correspondence $I_E=\beta F$ where $F$ is the free energy of the dual field theory and $\beta$ is the inverse black hole temperature.
In units $8 G=1$, the free energy reads
\begin{eqnarray}
 F&=&\left[2r\sqrt{f}\left(1+\frac{\Delta_\phi^-\phi^2}{4}\right)-rf'\sqrt{\frac{h}{f}}\right]\:\Big|^{r=\infty}-c g(r_0)\nn\\
&=& -\frac{\lambda}{2}+\sqrt{f_0}\phi_0\phi_1(2-\Delta_\phi)(1-\Delta_\phi)-c g(r_0).
\end{eqnarray}
Using the definition of the stress tensor in \cite{Balasubramanian:1999re}, the mass and the angular momentum of this black hole can be found to be
\begin{eqnarray}
&& M=2\sqrt{f}\left[r\left(1+\frac{\Delta_\phi^-\phi^2}{4}\right)-\sqrt{h}\right]\:\Big|^{r=\infty}=\frac{\lambda}{2}+\sqrt{f_0}\phi_0\phi_1(2-\Delta_\phi)(1-\Delta_\phi)\nn\\
&& J=c,
\end{eqnarray}
while the entropy and the angular velocity of the black hole are
\begin{equation}
 S=4\pi r_0,\qquad \Omega=-g(r_0).
\end{equation}
Evaluating the second equation in (\ref{g'}) on the outer horizon and at infinity one finds
\begin{equation}
 \left(r\sqrt{\frac{h}{f}}f'-2\sqrt{f h}\right)\:\Big|^{r=\infty}=\frac{S}{\beta}-2c g(r_0).
\end{equation}
Then, it is straightforward to check that the first law of thermodynamics holds,
\begin{equation}
 F=M-\frac{S}{\beta}-\Omega J.
\end{equation}

\subsubsection{Numerical results} 
\begin{figure}
\center{\epsfig{figure=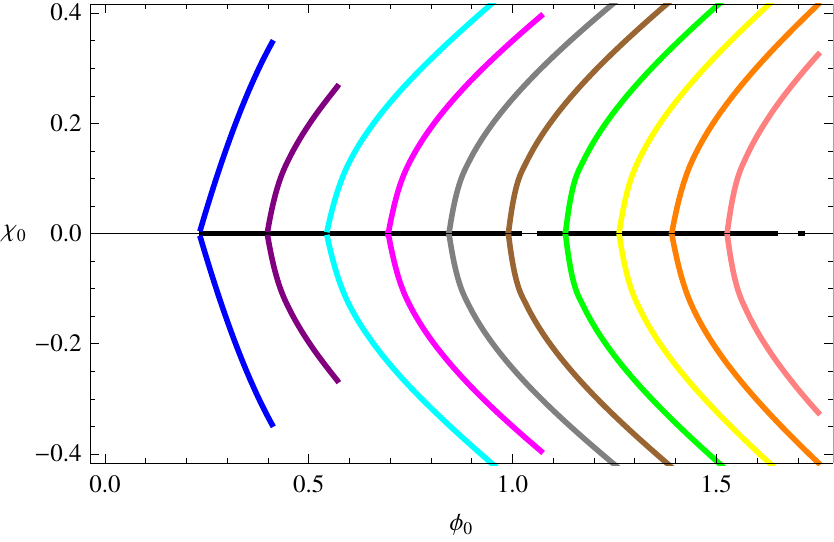, width=2.8in}\hspace{.6cm}\epsfig{figure=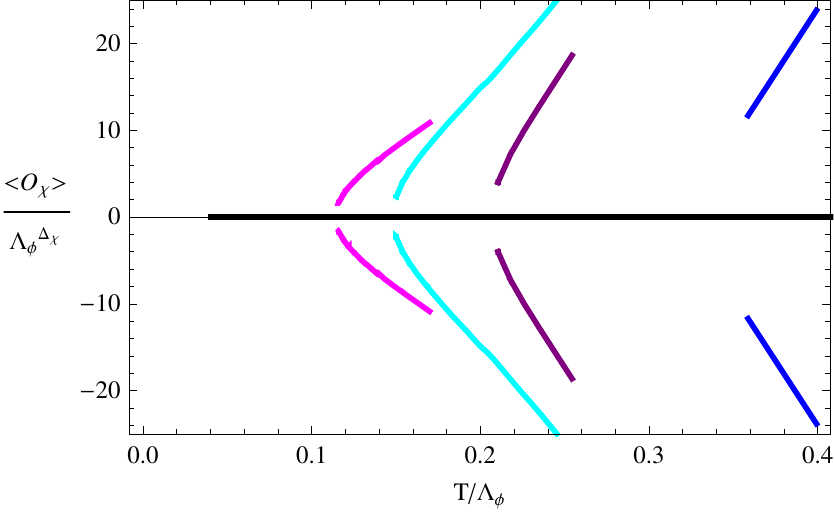,width=2.8in}}
\caption{
\label{freep}
a) Appearance of symmetry breaking phases with different nodes:
 from left to right the colors blue, purple, cyan, magenta, gray, brown, green, yellow, orange, pink and black represent phases of $n=0\hdots 9$ and the symmetric phase, respectively. $\chi_0$ and $\phi_0$ are the
values of these field on the black hole horizon.
b) The order parameter of the transition as a function of temperature for solutions with different nodes. From right to left, $n=0$ to $n=3$. }
\end{figure}

In this section we present numerical solutions to the equations of motion in (\ref{scalardyn}) and (\ref{backreac}) that describe phases with broken $\mathbb{Z}_2$ symmetry. 
Boundary conditions are set by choosing the value of $\phi_0=\phi(r_0)$ on the black hole horizon and the angular momentum
at infinity. We note that using symmetries, the $AdS_3$ radius and the location of horizon could be scaled away, $l=r_0=1$. The equations
of motion could be integrated as a boundary value problem using the shooting method. In the range numerically explored, the scalar $\phi(r)$ is monotically decreasing as a function of radius,
whereas $\chi(r)$ can acquire an arbitrary number of zeros at finite radii as we saw in the probe limit. For $\phi_0$ less than a critical value $\phi_c$, the $\chi$ field is zero everywhere in the bulk. At $\phi_0=\phi_c$ a new branch of solutions
is found with $\chi(r)$ monotically decreasing from the horizon to the boundary. Increasing $\phi_0$ further, at $\phi_c'$ another branch 
appears with one node in the wave function of $\chi$. As can be seen in figure \ref{freep}, the system admits higher node solutions for $\chi$ at higher $\phi_0$. For numerical purposes, we have chosen
$c=0.2$, $\lambda=8$ and $m_\chi^2=-m_\phi^2=0.84$. However, the qualitative features of the phase transition do not change as one varies
$c$ and $\lambda$. 

\begin{figure}[h]
\centerline{a) \epsfig{figure=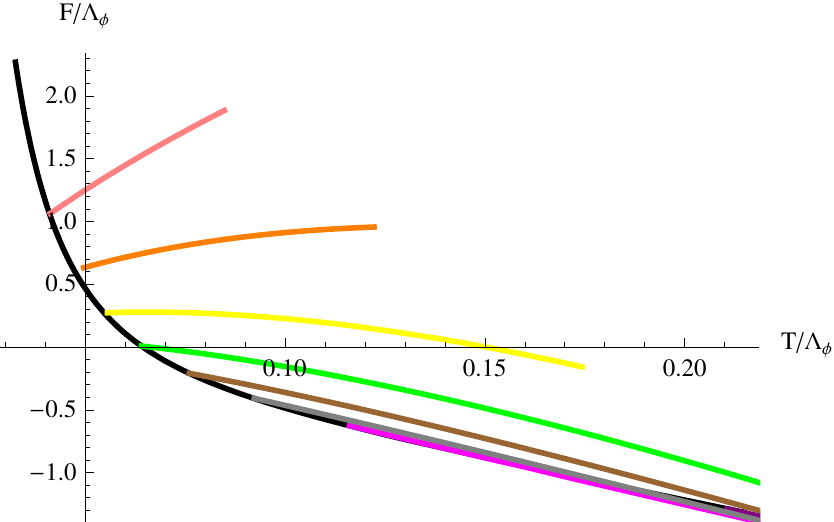, width=3in}  b) \epsfig{figure=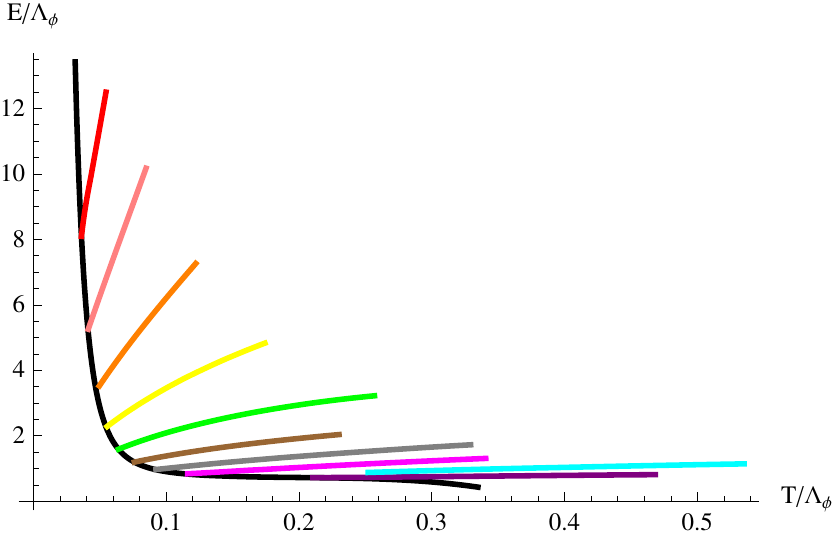, width=3in}}
\caption{
\label{freeenergy}
The thermodynamic properties of symmetry breaking phases. Colors represent same phases as in figure \ref{freep}. Higher $n$ have larger mass and free energy:
 a) the free energy  b) the mass of the black holes.}
\end{figure}

At finite temperature, there is a competition between the symmetric phase and broken phases with $n$ nodes. The numerical results suggest that
the symmetric phase always wins the competition by minimizing the free energy; see figure \ref{freeenergy}. At sufficiently large temperature, on top of the symmetric ground state
there is a tower of excited phases. Low $n$ phases are statistically more significant than high node ones. 
Decreasing the temperature gradually, at a critical temperature $T_1$ the $n=1$ phase joins the ground state and disappears from the spectrum. Higher $n$ phases join the ground state at lower
temperatures. This is a remarkable phenomenon, since intuitively one expects the order to be destroyed 
by high temperature fluctuations whereas these phases exist only in $T$ larger than some critical temperature \footnote{This phenomenon was observed in a similar 
system in four dimensions in \cite{Buchel:2009ge}. It was recently shown that the 4D high temperature ordered phases are perturbatively unstable \cite{Buchel:2010wk}. We thank Alex Buchel for bringing our attention to this reference.}.

\section{Continuous Symmetries}\label{cont}
\subsection{The Coleman-Mermin-Wagner theorem}
It was shown by Mermin-Wagner and Coleman independently that in 1+1 and 2+1 space-time dimensions at finite temperature, spontaneous continuous symmetry breaking is impossible \cite{Mermin:1966fe,Hohenberg:1967zz, Coleman:1973ci}.
This seems to be a no-go theorem for holographic superconductors in these dimensions. However, as was pointed out by Witten \cite{Witten:1978qu}, in the large $N$ limit, systems can have another phase in which
the symmetry is {\it almost} spontaneously broken and the fall off of correlation functions is power law. If this is the case, for finite $N$ one expects the $1/N$ corrections in AdS/CFT to
wash out the ordered phase and restore the symmetry in low temperature. In \cite{Anninos:2010sq}, by a one-loop calculation in $AdS_4$ black holes, it was shown that this is indeed the case.
There is no reason to believe that three dimensional gravity is different. In this section we will indeed see that there exist analogues of holographic superconductors in asymptotically $AdS_3$ space-times.
\subsection{The Abelian Higgs model}
The basic ingredients we need for a gravity dual to continuous symmetry breaking at finite temperature are black hole solutions charged under
a gauge field and a charged scalar field dual to the order operator in the field theory. 
The simplest theory of this form is the Abelian Higgs model. 
We will work in the probe limit where the back-reaction of the scalar and the gauge field on the geometry could be neglected. In an analogous way to previous sections, symmetry breaking occurs with the appearance
of a hairy solution that breaks the $U(1)$ gauge symmetry. The ansatz for the gauge field is $A^t=\phi(r)$ and 
$A^r$ and $A^\theta$ have been set to zero using the gauge transformations. 

The equations of motion of this system are
\begin{eqnarray}\label{eqmAbelian}
% &&f'+r\psi'^2+\frac{q^2re^f\phi^2\psi^2}{h^2}=0,\nn\\ 
% &&h'-\frac{f'}{2}h+\frac{r\phi'^2e^f}{2}-2r+\frac{m^2\psi^2 r}{2}=0\nn\\
&&\phi''+\frac{1}{r}\phi'-\frac{\psi^2}{h}\phi=0\nn\\
&&\psi''+\left(\frac{1}{r}+\frac{h'}{h}\right)\psi'-\left(\frac{m^2}{h}-\frac{\phi^2}{h^2}\right)\psi=0,
\end{eqnarray}
where $m$ and $q$ are the charge and the mass of the scalar field $\psi$ and $h(r)=g^{rr}=r^2-M$.

Although the gauge potential grows as $A\sim A_1+A_2 \log(r)$ near the boundary, the field strength goes to zero and asymptotically one recovers the AdS solution.
A thorough analysis of the boundary conditions in asymptotically $AdS_3$ spaces shows that the right boundary condition to impose is
to set $F^I=r \phi'(r)=const.$ on the boundary \cite{Marolf:2006nd,Maity:2009zz}. Another way to understand this is through Hodge duality. In the Hodge dual frame $d\phi=*F$ this is equivalent to fixing the slower falloff of the dual scalar field \cite{Marolf:2006nd}.
The fixed logarithmic term on the boundary will be interpreted as the chemical potential and the fluctuation $A_1$ represents the
expectation value of the boundary current dual to the gauge field.

At high temperature and fixed charge density, the only allowed solution is the charged BTZ discussed in the previous section.
As we lower the temperature, a branch of solutions appears with the scalar hair condensing near the horizon of the black hole and as we lower the temperature further, more and more solutions of this
type appear with $\rho=\langle \mathcal{J}^t\rangle$ acquiring more nodes at finite radii outside of the black hole horizon.

 Taking the value $m^2=-3/4$ one can expand the equations of motion near the boundary. The fall-off of the fields near the boundary is given by
\begin{eqnarray} 
&&\phi(r)=\phi_1+\phi_2\log(r)+\hdots\nn\\
&&\psi(r)=\frac{\psi_1}{\sqrt{r}}+\frac{\psi_2}{r^{3/2}}+\hdots.
\end{eqnarray}

Hence, the chemical potential is $\mu=-\phi_2\log(\epsilon)$ with $1/\epsilon$ representing the cut-off scale.
One numerical observation is that in the limit $T/T_c\to 0$ the condensate seems to go away. However for a complete zero temperature analysis it is essential 
to take into account the back-reaction of the condensate on the geometry \cite{Horowitz:2009ij}.

\begin{figure}
\center{\epsfig{figure=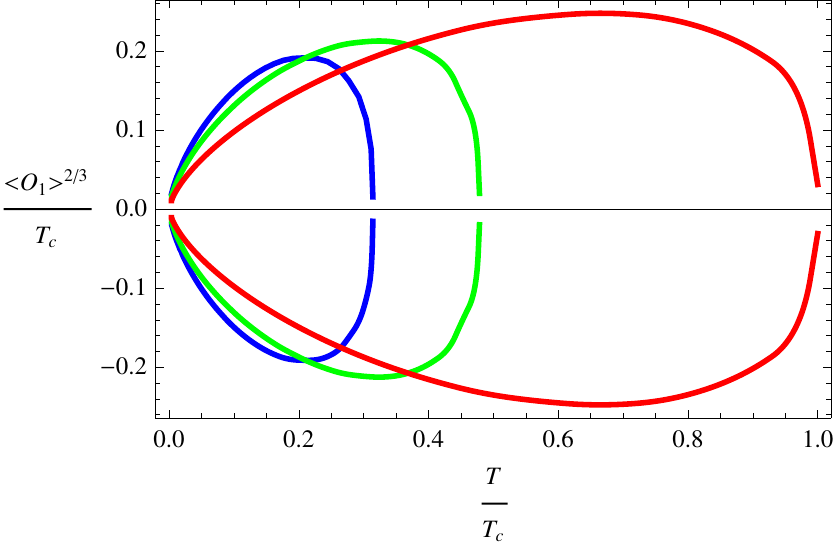, width=3in}}
\caption{The order parameter of this transition is the value of the condensate at fixed charged density. The above plot shows the temperature dependence of the order parameter
on solutions with different number of nodes. From right to left red, green and blue respectively represent the zero, one and two node solutions.
\label{fp}
}
\end{figure}

\section{Discussion}
In this paper we discussed examples of symmetry breaking in 3D gravity that could be thought of as phases of some strongly coupled 2D
conformal field theories with large central charges. In the case of a discrete broken symmetry, we observed a rich phase space of solutions with high temperature order.
In condensed matter, there are known systems with similar exotic thermodynamics \cite{Chubukov:1992ei}. It would be extremely interesting to investigate this connection explicitly.

If there exist theories of quantum gravity with small values of AdS radius in planck units, one expects them to be dual to 2D field theories with small central charge. An interesting implication of the Coleman-Mermin-Wagner theorem is that it seems such theories of gravity can never admit continuous phase transitions at zero or any finite temperature. 

 There are other remaining interesting issues one can investigate. Firstly, it would be a natural step to try to embed these models in string theory. Holographic symmetry breaking in 1+1D in the context of D1-D5 CFT was previously studied in \cite{Skenderis:2006ah}. 
Secondly, the behavior we observe in the zero temperature limit of holographic superconductors in three dimensions seems curious. As was mentioned in the previous section, the numerical results in the probe regime suggest that at $T\to 0$ the continuous symmetry is again
restored. For low enough temperatures the value of the condensate increases with temperature. 

We would like to especially thank Alejandra Castro and Omid Saremi for extremely valuable discussions without which this work would not have been completed. 
We are also indebted to Alexander Maloney and Aaron C. Vincent for a critical reading of the draft.

% 
% \appendix
% 
% \section{appendix name}
% 

\end{document}